# Evidence of gap anisotropy in superconducting $YNi_2B_2C$ using directional point contact spectroscopy

P Raychaudhuri[a], D Jaiswal-Nagar, Goutam Sheet[b], and S Ramakrishnan

Department of Condensed Matter Physics and Materials Science, Tata Institute of Fundamental Research, Homi Bhabha Road, Colaba, Mumbai 400005, India

and

H Takeya

National Institute for Materials Science, 3-13 Sakura, Tsukuba, Ibaraki 305-0003, Japan

We present a study of the anisotropy in the superconducting energy gap in a single crystal of $YNi_2B_2C$ ($T_c \sim 14.6K$) using directional point contact spectroscopy. The superconducting energy gap at 2.7K, when measured for $I \| c$, is 4.5 times larger than that for $I \| a$. The energy gaps in the two directions also have different temperature dependences. Our results support a scenario with $s+g$ like symmetry.

73.23.Ad, 74.80.Fp, 74.70.-b, 74.00.00

[a] Corresponding author, E-mail: pratap@tifr.res.in

[b] Corresponding author, E-mail: goutam@tifr.res.in

Determination of the symmetry of the order parameter is a central step in understanding the pairing mechanism of any superconducting system. Conventional superconductors, i.e. elemental superconductors and a vast majority of their alloys, are characterized by an isotropic energy gap with s-wave pairing symmetry. The interest in the pairing symmetry has revived with the discovery of a new class of superconductors, where the pairing is highly anisotropic with the gap function going to zero for certain $k$ directions[1]. These are the high $T_c$ cuprates[2], the triplet superconductor[3] $Sr_2RuO_4$, organic superconductors[4,5] and several heavy Fermion compounds[6,7]. Since a zero in the gap function can arise only from a non-trivial representation of the symmetry group, such gap structures point towards unconventional pairing interactions mediating the superconductivity. Superconductivity in these materials is widely believed to originate from interactions of purely electronic origin[1].

In this work, we focus our attention on one member of another class of novel materials, namely, the quaternary borocarbide superconductor, $YNi_2B_2C$. Ten years after superconductivity was reported in this compound[8,9], the symmetry of the order parameter in $YNi_2B_2C$ has still remained an outstanding issue. $YNi_2B_2C$ has a tetragonal crystal structure with $a$=3.526Å and $c$=10.543Å. Early studies on this compound suggested a s-wave symmetry[10,11,12] mediated by conventional electron-phonon interaction. However, in recent years several experimental studies such as thermal conductivity[13,14], specific heat[15], tunneling spectroscopy[16] and photoemission spectroscopy[17] have not only provided evidence of a large anisotropy but also the existence of nodes in the superconducting gap function. Furthermore, the angular variation of the various physical properties such as thermal conductivity[13], critical fields etc. shows a robust 4-fold anisotropy in the *a-b* plane in the superconducting state, while the normal state properties remain almost isotropic[18]. These results provide indirect evidence of a strong anisotropy in the superconducting order parameter, and have been interpreted based on either anisotropic *s*-wave (s+g wave)[19] with point nodes or *d*-wave symmetry[20] with line nodes. Considering that d and s+g wave symmetries are likely to have their origins from very different pairing interactions, more direct information regarding the gap anisotropy in $YNi_2B_2C$ is highly desirable.

Directional point contact (DPC) spectroscopy, i.e. where conductance spectra (dI/dV versus V) are recorded by injecting current through a ballistic point contact along different crystallographic directions in the superconductor, is a powerful tool to investigate the gap anisotropy in unconventional superconductors[21]. This technique allows a direct determination of the gap and its temperature dependence along various directions in the crystal. In this paper, we report the DPC measurements in a high quality single crystal of $YNi_2B_2C$. The spectra were recorded by injecting the current either along *c* or along *a,* thereby measuring the gap along these two directions in the temperature range 2.6K-15K. Since the structure of the gap function for s+g symmetry and d-wave symmetry would be different along these two directions, the above measurements allow us to distinguish between these two symmetries. Our results provide evidence of sharp minima in the gap structure of $YNi_2B_2C$ in the basal plane, consistent with the s+g-wave scenario. Furthermore, normalized energy gap along different crystallographic directions have different temperature dependences suggesting that the amplitude of s and g in the order parameter have different temperature dependence.

The YNi$_2$B$_2$C single crystal was grown by traveling solvent floating zone method using an image furnace. The crystal was then cut into a rectangular parallelepiped of size 0.5mm×0.5mm×2mm with the long axis along *a*. This allowed us to have relatively large facets on the [100] and [001] planes. The T$_c$~14.6K was determined from ac susceptibility measured at 15kHz (Fig.1). Under the application of a dc magnetic field, H||*c*, the signature of a "peak effect", i.e. a dip in the real part of the ac susceptibility ($\chi'$) associated with an order-disorder transformation of the vortex lattice, was observed down to 1000 Oe. Since the inter-vortex spacing for this field (~1500Å) is larger than the penetration depth[22] ($\lambda$~1000Å), this points towards the very low defect density in the crystal. The crystal quality was further confirmed from the observation of dHvA oscillations. Resistivity measurement on similar crystals showed the residual resistivity to be ~4$\mu\Omega$-cm, corresponding to a mean free path ~110Å at low temperatures[23]. The crystal facets were polished with fine emery paper followed by 5$\mu$m size alumina powder to obtain a mirror-finished surface prior to the point contact measurement. For the point contact measurement, a mechanically cut fine tip made of 0.25mm thick gold wire was brought in contact with the [100] or [001] facet (for I||*a* and I||*c* respectively) of the crystal using a 100 threads per inch differential screw arrangement in a liquid He cryostat. The temperature stability was better than 10mK in temperature range of our measurement, i.e. 2.6K-15K. To establish a ballistic contact the tip was first engaged on the sample and then gradually withdrawn in small steps till a ballistic contact was established. Details of this method have been described elsewhere[24]. For all the spectra reported in this paper the point contact resistance ($R_N$) in the normal state was in the range 10-20$\Omega$, from which the point contact diameter (*d*) was estimated to be[25],

$$d = \left\{ \left(\frac{4}{3\pi}\right)\left(\frac{\rho l}{R_N}\right) \right\} \sim 30\text{-}40\text{Å}.$$ Therefore our measurements were well into the ballistic limit (*d*<*l*) of the point contact. A four-probe (constant current) modulation technique operating at 372 Hz was used to directly measure the differential resistance ($R_d$=(dV/dI)) as a function of voltage from which the differential conductance ($G=1/R_d$) was calculated.

Figures 2(a) and 2(b) show two sets of point contact spectra for I||*a* and I||*c*, respectively measured in the temperature range 2.6K to 15K. A gap related feature, is clearly discernible in the spectra in both directions at the lowest temperature. However, even a visual inspection shows that the gap value is much larger for I||*c* than for I||*a*. For I||*c*, the gap feature in the spectra can be observed up to 14K. However, for the "small gap" direction I||*a*, no feature was resolved in the conductance curve for T>7K. The conductance curves were fitted with the ususal Blonder-Tinkham-Klapwijk model[26] (solid lines) using the superconducting energy gap $\Delta$, the barrier height coefficient Z and the broadening parameter[27] $\Gamma$ as fitting parameters. In addition to the kind of spectra shown in Fig. 2 for some contacts on the [010] facet, we encountered some spectra, where the fitted gap value was intermediate between these two values. These contacts

were however very unstable and slight modification of the contact by mechanical vibration or while trying to heat the sample above the base temperature resulted in a spectrum such as the one shown in figure 2(b). We believe that these spectra resulted due to the current not being injected perpendicular to the [010] facet due to surface roughness.

We first concentrate on the G-V curves at the lowest temperatures (2.7K). From the best fit parameters, the superconducting gaps in the two directions were determined to be $\Delta_{I\|c}$=1.8±0.1meV and $\Delta_{I\|a}$ =0.415±0.08meV, respectively. This corresponds to a gap anisotropy of $\Delta_{I\|c}/\Delta_{I\|a}$~4.5. However, it is important to note that for a ballistic contact, when I is injected along a particular direction **n,** the current results from an average over the entire Fermi surface though dominant contribution comes from Fermi surface regions close to $k_{Fn}$, while the contribution from Fermi surface regions perpendicular to **n** is zero[28]. Therefore, the maximum ($\Delta_{max}$) and the minimum ($\Delta_{min}$) values of the gap are actually larger and smaller than $\Delta_{I\|c}$ and $\Delta_{I\|a}$ respectively. Thus $\Delta_{I\|c}/\Delta_{I\|a}$ gives a lower bound of the gap anisotropy on the Fermi surface. The relative broadening of the spectra, characterized by $\Gamma/\Delta$ in the two directions is also different: $(\Gamma/\Delta)$~0.32 for I∥c and $(\Gamma/\Delta)$~0.77 for I∥a.

We can now compare these results with the two order parameter symmetries proposed for this compound, namely, d and s+g [19,20]. The gap functions corresponding to these two symmetries for a 3 dimensional Fermi surface in polar co-ordinates are $\Delta(\mathbf{k})= \Delta_0[\sin^2\theta\sin(2\phi)]$ and $\Delta(\mathbf{k})= (\Delta_0/2)[1-\sin^4\theta\cos(4\phi)]$, where the convention for $\theta$ and $\phi$ are the same as in Ref.13. These two symmetries are schematically shown in figure 2. For the d-wave symmetry the gap function has line nodes perpendicular to the basal plane extending to the poles. Therefore, the gap function has nodes along both [100] and [001] directions. For s+g symmetry (with equal amplitude of s and g) on the other hand, the gap is fully formed close to the poles and has point nodes only along [100] and [010]. Though our point contact measurement cannot unambiguously identify a node in the gap function due to Fermi surface averaging described before, the large value of $\Delta_{I\|c}/\Delta_{I\|a}$ observed at low T suggests that the gap function has sharp minima in the basal plane and a fully gapped nature close to the poles. This is clearly consistent with the s+g symmetry.

We now look at the anisotropy in $\Gamma/\Delta$ in the two directions. Though $\Gamma$ is commonly attributed to the quasiparticle lifetime limited broadening of the spectra, in a superconductor with an anisotropic gap function, a broadening in the point contact spectrum is also expected to arise from the averaging over the superconducting energy gap. The latter, though physically distinct in origin from the former, is indistinguishable within experimental errors. To illustrate this point, we have simulated a set of spectra using a distribution of superconducting energy gaps, keeping the mean value of the gap $<\Delta>=0.6$meV and setting $\Gamma=0$. The G-V curves are computed from

$I = \sum_i I_{BTK}(Z = 0.4, T, \Delta_i, \Gamma = 0, V)$, where $I_{BTK}$ is the well-known BTK expression for

the current[26] and the sum runs over the distribution of gap values. Four G-V curves (normalized with respect to their conductance values at high bias) with varying distribution of gap values are shown in Figure 3. All the curves could be fitted (solid lines) using a single BTK function when the broadening parameter $\Gamma$ is used as an additional fitting parameter. The deviation of the fitted curve from the original curve is smaller than our level of experimental accuracy. Furthermore, the fitted value of $\Gamma/\Delta$ increases monotonically with the standard deviation of the distribution over $\Delta$, i.e.

$\sigma(\Delta) = \sqrt{\sum_i (\Delta_i - \langle \Delta \rangle)^2}$, showing that $\Gamma/\Delta$ is a measure of the variation is

superconducting energy gap values (see *inset* Figure 3). Therefore the larger value of $(\Gamma/\Delta)_{I\|a}$ compared to $(\Gamma/\Delta)_{I\|c}$ in our measurements suggests that the superconducting energy has a larger variation close to the poles than in the basal plane along [001]. Since in the s+g symmetry the gap function is "flat" close to the poles but has a sharp variation from zero close to the point nodes along [100] the larger $\Gamma/\Delta$ value for $I\|a$ further supports s+g scenario[29,30].

Finally, we concentrate on the temperature dependence of the superconducting energy gap. Figure 4 shows the temperature variation for $\Delta_{I\|c}$ and $\Delta_{I\|a}$ extracted from the BTK fits of the spectra at various temperatures. There is a clear deviation from the expected temperature variation for an isotropic BCS superconductor for both $I\|a$ and $I\|c$ with the gap decreasing faster than the BCS prediction. This is consistent with the existence of gap zeros[21]. However, the striking observation is difference in the temperature dependence of the normalized energy gap in the two directions (inset of figure 4). For I∥c the gap persists up to 14K. For I∥a the gap decreases rapidly from its low temperature value and no gap can be resolved at temperatures >7K. Even taking into account the fact that a small gap will get smeared due to thermal broadening at higher temperatures the rapid decrease in $\Delta_{I\|a}$ from the low temperature value is beyond any experimental error. This cannot be explained for a gap function of the form $\Delta(\mathbf{k})=\Delta_0 f(\mathbf{k})$ where the temperature dependence comes from $\Delta_0$ alone, and therefore should be same for all $\mathbf{k}$. It has to be however kept in mind that in the s+g model the amplitude of the s and g are fine tuned to be equal to obtain point nodes along [100] and [010] directions. The equality in amplitude of inequivalent representations like s and g however seems "accidental", with no symmetry reason why they should even be close[19]. Experimentally on the other hand the ratio of $\Delta_{max}/\Delta_{min}$ has been estimated to be between 10-100 from different experiments[13,14]. It is therefore possible that the s and g amplitude following different temperature dependences with the gap zero being present only at low temperatures. This would give rise to a gap function of the form $\Delta(\mathbf{k})=\Delta_{0s}+\Delta_{0g}g(\mathbf{k})$ with different temperature dependence[31,32] of $\Delta_{0s}$ and $\Delta_{0g}$. In this situation the shape of the gap function would be temperature dependent giving rise to different temperature dependence of the gap along different directions.

In conclusion, directional point contact spectroscopy in $YNi_2B_2C$ reveals a pronounced anisotropy in the superconducting energy gap with a fully gapped structure along the c-direction and sharp minima in the basal plane. This rules out the possibility of d-wave symmetry and is consistent with the s+g scenario proposed in this system. However, the normalized superconducting gap have different temperature dependence for I||a and I||c suggesting that the ratio of the amplitude of s and g component in the order parameter vary with temperature.

## Acknowledgements

We would like to thank Drs. N Trivedi, M. Randeria and A. Paramenkanti for illuminating discussions and Prof. A K Grover for critically reading the manuscript. We would like to thank Dr T. Terashima for sharing his results on dHvA measurements on this crystal. Two of us (GS and DS) would like to thank the TIFR Endowment Fund for partial financial support.

**1) Temperature dependence of ac susceptibility (measured at 15kHz) for $YNi_2B_2C$ at different magnetic fields. $T_c$ at zero field is 14.5 K. Signature of peak effect (shown by arrows) is observed down to 1000 Oe.**

**2) Point contact Andreev reflection spectra at different temperatures for (a) I||a and (b) I||c. Open circles are experimental data and solid lines are BTK fits to the spectra. The conductance curves are normalized by their respective values at high bias. (c) and (d) show the gap functions corresponding to s+g and d wave symmetry**

respectively where the radial distance from the origin is proportional to the magnitude of the gap in that *k* direction.

3) Simulated spectra at 2.7K (open circles) assuming different distribution of Δ. For (1) $\Delta_i$=0.6. For (2) $\Delta_i$=0.4, 0.6, 0.8 meV, for (3) $\Delta_i$=0.2, 0.4, 0.6, 0.8, 1.0meV and for (4) $\Delta_i$=0.0, 0.2, 0.4, 0.6, 0.8, 1.0, 1.2meV respectively. Simulated curves are fitted (solid lines) with BTK theory with broadening parameter Γ included. The best-fit parameters for Δ and Γ are also shown in the figure. *inset :* evolution of Γ/Δ with σ(Δ).

4) Temperature dependence of $\Delta_{I \parallel c}$ (solid circles) and $\Delta_{I \parallel a}$ (open boxes). Solid lines show the expected temperature variation from BCS theory for an isotropic gap. *inset:* Temperature variation of the normalized energy gap for I∥c and I∥a. Δ is normalized to its value at the lowest temperature. The solid lines are guides to the eye.

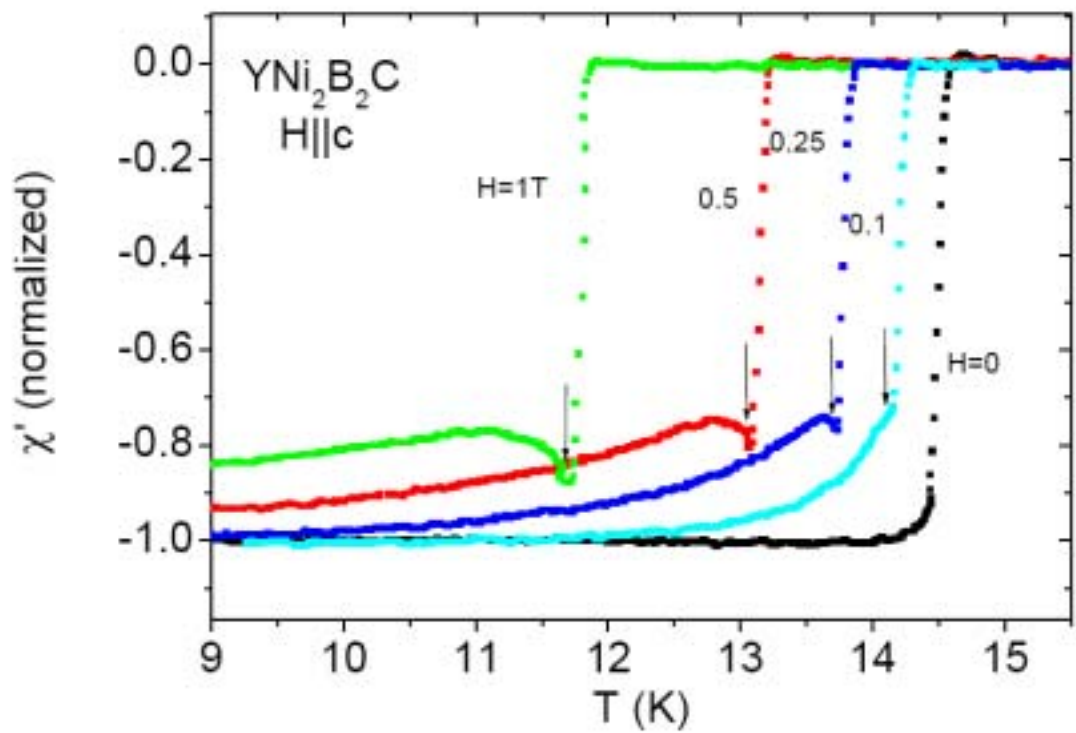

Figure 1

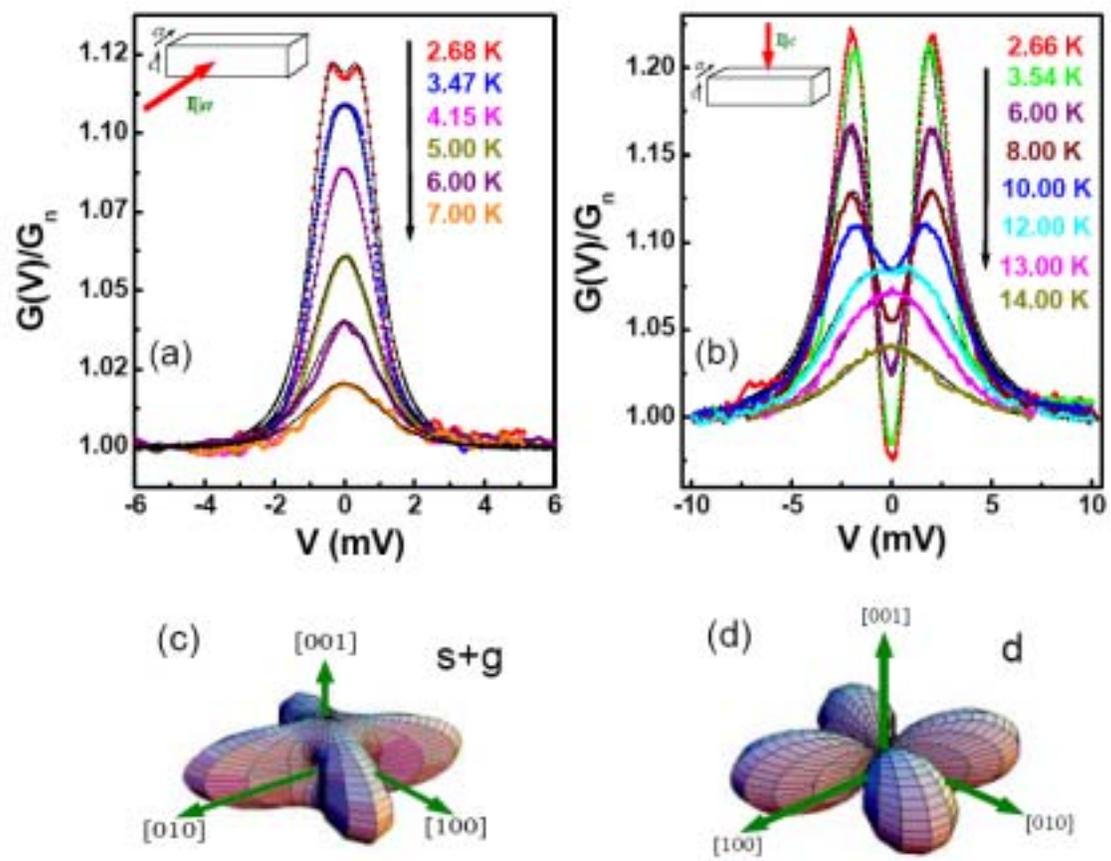

Figure 2

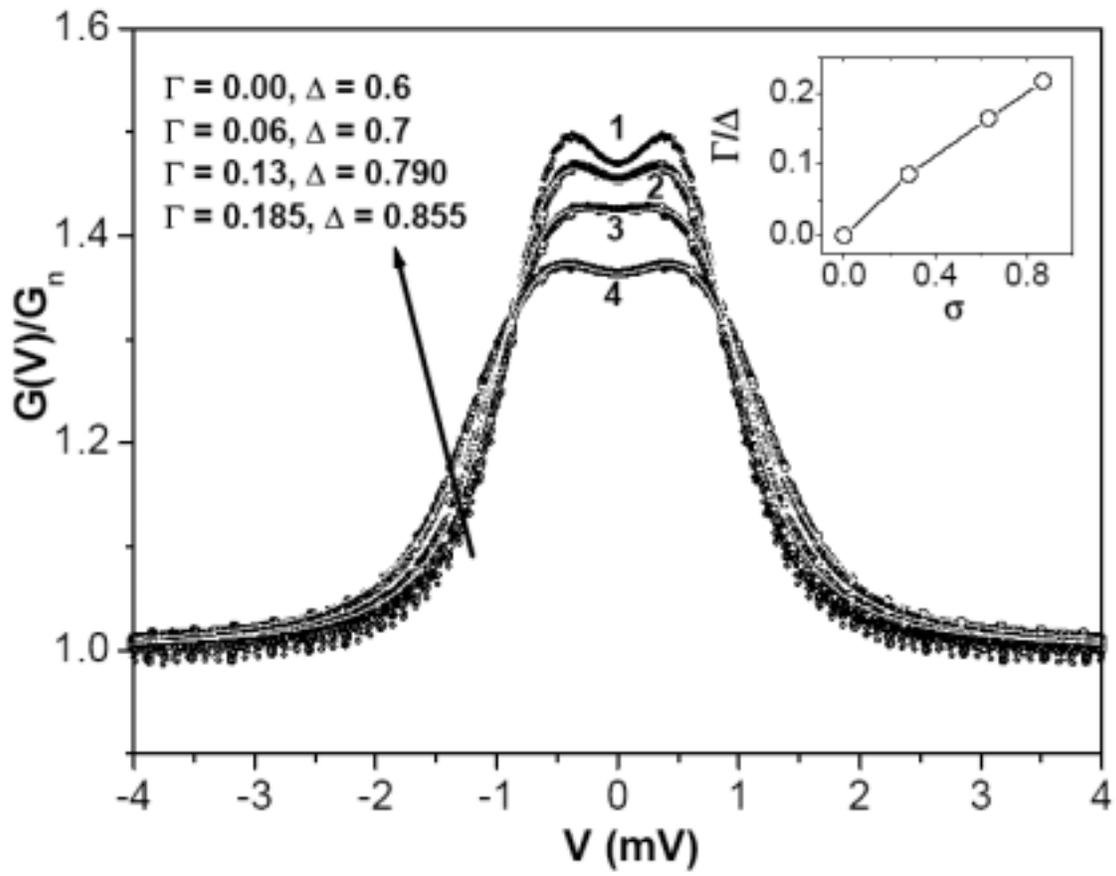

Figure 3

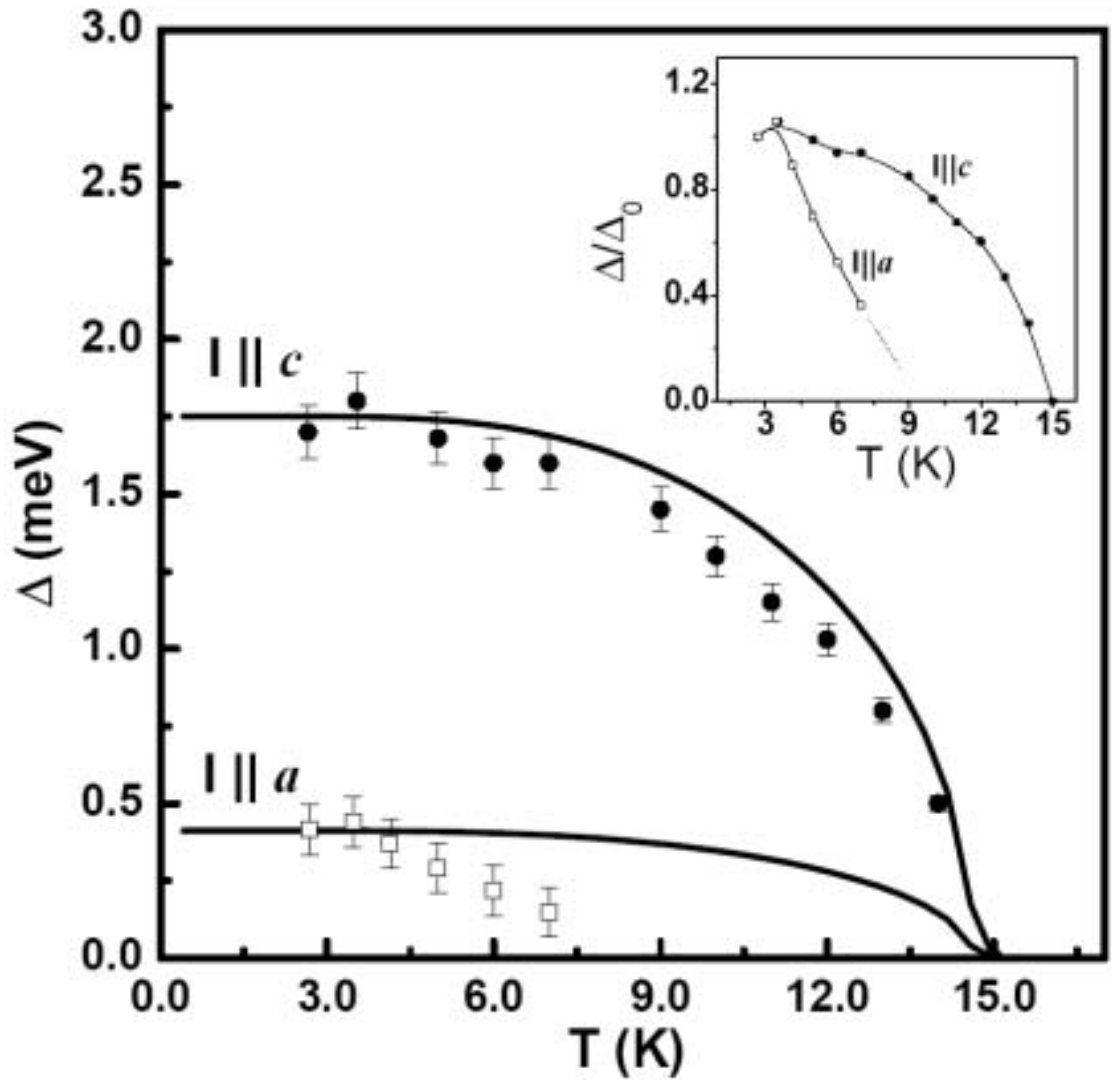

Figure 4

---

[29] Further evidence of s+g symmetry can be obtained from the evolution of the gap with the addition of impurities. For s+g symmetry the gap function should become more isotropic with the addition of impurities. This is consistent with the photoemission results in ref. 17. In contrast for a d-wave superconductor the gap should vanish with the addition of impurities.

[30] In principle, different values of the broadening in different directions can also arise from the anisotropy of the Fermi surface. However since the normal state properties of $YNi_2B_2C$ are nearly isotropic suggesting a nearly spherical Fermi surface (ref.18), the anisotropy in the Fermi surface is unlikely to have a pronounced effect in the spectra

[31] Q Yuan and P Thalmeier, Phys. Rev. B **68**, 174501 (2003).

[32] This possibility has been theoretically explored in ref. 31 using BCS theory. The authors indeed observe that a state with equal amplitudes of s and g at T=0 evolves with temperature and the amplitudes do not remain equal. Their model calculations however show that the deviation is of the order of 1% close to $T_c$.